\documentclass[aps,prl,twocolumn,superscriptaddress,amsfont,graphicx,nofootinbib,preprintnumbers]{revtex4}%
\usepackage{color,graphicx,epsfig}
\usepackage{ifpdf}
\usepackage{amsmath}
\usepackage{bm}
\usepackage{color}
\usepackage[english]{babel}
\usepackage{graphicx}%
\usepackage{amsfonts}%
\usepackage{amssymb}
\usepackage{braket}
\usepackage{hyperref}
\usepackage{enumerate}

\definecolor{nicered}{rgb}{0.7,0.1,0.1}
\definecolor{nicegreen}{rgb}{0.1,0.5,0.1}
\hypersetup{colorlinks,citecolor= nicegreen,linkcolor= nicered}

\newcommand{\beq}{\begin{equation}}
\newcommand{\eeq}{\end{equation}}
\newcommand{\beqa}{\begin{eqnarray}}
\newcommand{\eeqa}{\end{eqnarray}}

\definecolor{Red}{rgb}{1.,0.,0.}

\newcommand{\rd}{{\rm d}}
\newcommand{\LO}{\rm LO}
\newcommand{\NLO}{\rm NLO}
\newcommand{\NNLO}{\rm NNLO}
\newcommand{\ds}{{\rm d}\hat{\sigma}}
\newcommand{\nn}{\nonumber}

\def\tauone{{\cal T}_1}
\def\tauncut{{\cal T}_N^{cut}}
\def\tauonecut{{\cal T}_1^{cut}}

\def\mysection#1{{{\bf #1}.~}}
\def\OMIT#1{}

\preprint{NSF-KITP-16-054}

\begin{document}

\def\Maryland{Maryland Center for Fundamental Physics, University of Maryland, College Park, Maryland 20742, USA}
\def\Argonne{High Energy Physics Division, Argonne National Laboratory, Argonne, IL 60439, USA}
\def\Northwestern{Department of Physics \& Astronomy, Northwestern University, Evanston, IL 60208, USA}

\title{Single-inclusive jet production in electron-nucleon collisions through next-to-next-to-leading order in perturbative QCD}

\author{Gabriel Abelof}     
\email[Electronic address:]{gabelof@anl.gov}
\affiliation{\Argonne}
\affiliation{\Northwestern}

\author{Radja Boughezal}     
\email[Electronic address:]{rboughezal@anl.gov}
\affiliation{\Argonne}

\author{Xiaohui Liu}
\email[Electronic address:]{xhliu@umd.edu}
\affiliation{\Maryland}

\author{Frank Petriello}     
\email[Electronic address:]{f-petriello@northwestern.edu}
\affiliation{\Argonne}
\affiliation{\Northwestern}

\date{\today}
\begin{abstract}
We compute the ${\cal O}(\alpha^2\alpha_s^2)$ perturbative corrections to inclusive jet production in electron-nucleon collisions.  This process is of particular interest to the physics program of a future Electron Ion Collider (EIC).  We include all relevant partonic processes, including deep-inelastic scattering contributions, photon-initiated corrections, and parton-parton scattering terms that first appear at this order.  Upon integration over the final-state hadronic phase space we validate our results for the deep-inelastic corrections against the known next-to-next-to-leading order (NNLO) structure functions.  Our calculation uses the $N$-jettiness subtraction scheme for performing higher-order computations, and allows for a completely differential description of the deep-inelastic scattering process.  We describe the application of this method to inclusive jet production in detail, and present phenomenological results for the proposed EIC.  The NNLO corrections have a non-trivial dependence on the jet kinematics and arise from an intricate interplay between all contributing partonic channels.

\end{abstract}

\maketitle

\section{Introduction} \label{sec:intro}

The production of hadrons and jets at a future Electron Ion Collider (EIC) will play a central role in understanding the structure of the protons and nuclei which comprise the visible matter in the universe.  Measurements of inclusive jet and hadron production with transversely polarized protons probe novel phenomena within the proton such as the Sivers function~\cite{Kang:2011jw}, and address fundamental questions concerning the validity of QCD factorization.  Event shapes in jet production can give insight into the nuclear medium and its effect on particle propagation~\cite{Kang:2012zr}.  The precision study of these processes at a future EIC will provide a much sharper image of proton and nucleus structure than is currently available.  Progress is needed on both the experimental and theoretical fronts to achieve this goal.  Currently, much of our knowledge of proton spin phenomena, such as the global fit to helicity-dependent structure functions~\cite{deFlorian:2008mr}, comes from comparison to predictions at the next-to-leading order (NLO) in the strong coupling constant.  Theoretical predictions at the NLO level for jet and hadron production in DIS suffer from large theoretical uncertainties from uncalculated higher-order QCD corrections~\cite{Hinderer:2015hra} that will eventually hinder the precision determination of proton structure.  In some cases even NLO is unknown, and an LO analysis fails to describe the available data~\cite{Gamberg:2014eia}.  Given the high luminosity and expected precision possible with an EIC, it is desirable to extend the theoretical precision beyond what is currently available.  For many observables, a prediction to next-to-next-to-leading order (NNLO) in the perturbative QCD expansion will ultimately be needed.

An important step toward improving the achievable precision for jet production in electron-nucleon collisions was taken in Ref.~\cite{Hinderer:2015hra}, where the full NLO ${\cal O}(\alpha^2\alpha_s)$ corrections to unpolarized $lN \to jX$ and $lN \to hX$ scattering were obtained.  Focusing on single-inclusive jet production for this discussion, it was pointed out that two distinct processes contribute: the deep-inelastic scattering (DIS) process $lN \to ljX$, where the final-state lepton is resolved, and $\gamma N \to jX$, where the initial photon is almost on-shell and the final-state lepton is emitted collinear to the initial-state beam direction.  Both processes were found to contribute for expected EIC parameters, and the shift of the leading-order prediction was found to be both large and dependent on the final-state jet kinematics.

Our goal in this manuscript is to present the full ${\cal O}(\alpha^2\alpha_s^2)$ NNLO contributions to single-inclusive jet production in electron-nucleon collisions, including all the relevant partonic processes discussed above.  Achieving NNLO precision for jet and hadron production is a formidable task.  The relevant Feynman diagrams which give rise to the NNLO corrections consist of two-loop virtual corrections, one-loop real-emission diagrams, and double-real emission contributions.  Since these three pieces are separately infrared divergent, some way of regularizing and canceling these divergences must be found.  However, theoretical techniques for achieving this cancellation in the presence of final-state jets have seen great recent progress.  The introduction of the $N$-jettiness subtraction scheme for higher order QCD calculations~\cite{Boughezal:2015dva,Gaunt:2015pea} has lead to the first complete NNLO descriptions of jet production processes in hadronic collisions.  During the past year several NNLO predictions for processes with final-state jets have become available due to this theoretical advance~\cite{Boughezal:2015dva,Boughezal:2015aha,Boughezal:2015ded,Boughezal:2016dtm,Boughezal:2016isb,Boughezal:2016yfp}.  In some cases the NNLO corrections were critical in explaining the observed data~\cite{Boughezal:2016yfp}.  We discuss here the application of the $N$-jettiness subtraction scheme to inclusive jet production in electron-proton collisions.   Our result includes both the DIS and photon-initiated contributions, and allows arbitrary selection cuts to be imposed on the final state.  Upon integration of the DIS terms over the final-state hadronic phase space we compare our result against the known NNLO prediction for the inclusive structure function, and we find complete agreement.  We present phenomenological results for proposed EIC parameters.  We find that all partonic channels, including new ones that first appear at this order, contribute in a non-trivial way to give the complete NNLO correction.  We note that the NNLO corrections to similar processes, massive charm-quark production in deeply inelastic scattering and dijet production, were recently obtained~\cite{Berger:2016inr,Currie:2016ytq}.


\section{Lower-order results}\label{sec:low}

We begin by discussing our notation for the hadronic and partonic cross sections, and outlining the expressions for the LO and NLO cross sections.  We will express the hadronic cross section in the following notation:
\begin{equation}
\rd\sigma = \rd\sigma_{\text{LO}}+\rd\sigma_{\NLO}+\rd\sigma_{\NNLO}+\ldots \,,
\end{equation}
where the ellipsis denotes neglected higher-order terms.  The LO subscript refers to the ${\cal O}(\alpha^2)$ term, the NLO subscript denotes the ${\cal O}(\alpha^2\alpha_s)$ correction, while the NNLO subscript indicates the ${\cal O}(\alpha^2\alpha_s^2)$ contribution.  For the partonic cross sections, we introduce superscripts that denote the powers of both $\alpha$ and $\alpha_s$ that appear.  For example, the leading quark-lepton scattering process is expanded as 
\begin{equation}
\rd\hat{\sigma}_{ql}= \rd\hat{\sigma}_{ql}^{(2,0)}+\rd\hat{\sigma}_{ql}^{(2,1)}+\rd\hat{\sigma}_{ql}^{(2,2)}+\ldots \,.
\end{equation}
Here, the $\rd\hat{\sigma}_{ql}^{(2,0)}$ denotes the ${\cal O}(\alpha^2)$ correction, while $\rd\hat{\sigma}_{ql}^{(2,1)}$ indicates the  ${\cal O}(\alpha^2\alpha_s)$ term.

\begin{figure}[h]
\centering
\includegraphics[width=1.5in]{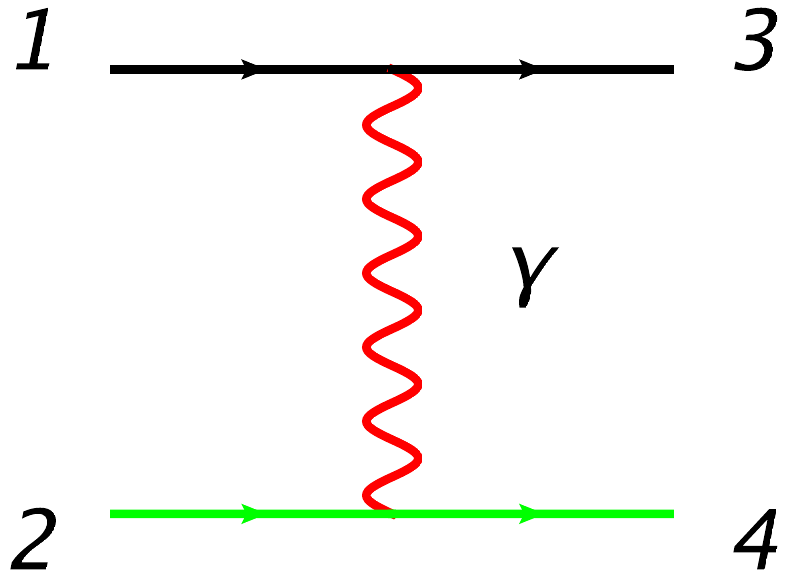}%
\caption{Feynman diagram for the leading-order process $q(p_1)+l(p_2) \to q_(p_3)+l(p_4)$.  We have colored the photon line red, the lepton lines green and the quark lines black.} \label{fig:LOdiag}
\end{figure}

 The leading-order hadronic cross section can be written as a convolution of parton distribution functions with a partonic cross section, 
\begin{eqnarray}
\label{eq:sigLO}
\rd\sigma_{\text{LO}} &=& \int \frac{\rd \xi_1}{\xi_1} \frac{\rd \xi_2}{\xi_2} \sum_q \left[ f_{q/H}(\xi_1) f_{l/l}(\xi_2) \rd\hat{\sigma}_{ql}^{(2,0)} \right. \\ \nonumber
	&+& \left. f_{\bar{q}/H}(\xi_1) f_{l/l}(\xi_2) \rd\hat{\sigma}_{\bar{q}l}^{(2,0)}\right]. 
\end{eqnarray} 
Here, $f_{q/H}(\xi_1)$ is the usual parton distribution function that describes the distributions of a quark $q$ in the hadron $H$ carrying a fraction $\xi_1$ of the hadron momentum.  $f_{l/l}(\xi_2)$ is the distribution for finding a lepton with momentum fraction $\xi_2$ inside the original lepton.  At leading order this is just $f_{l/l}(\xi_2)=\delta(1-\xi_2)$, but it is modified at higher orders in the electromagnetic coupling by photon emission.  
$d\hat{\sigma}_{ql}^{(2,0)}$ is the differential partonic cross section.  At leading order only the partonic channel $q(p_1)+l(p_2) \to q(p_3)+l(p_4)$ and the same process with anti-quarks instead contribute.  The relevant Feynman diagram is shown in Fig.~\ref{fig:LOdiag}. It is straightforward to obtain these terms.

\begin{figure}[h]
\centering
\includegraphics[width=3.0in]{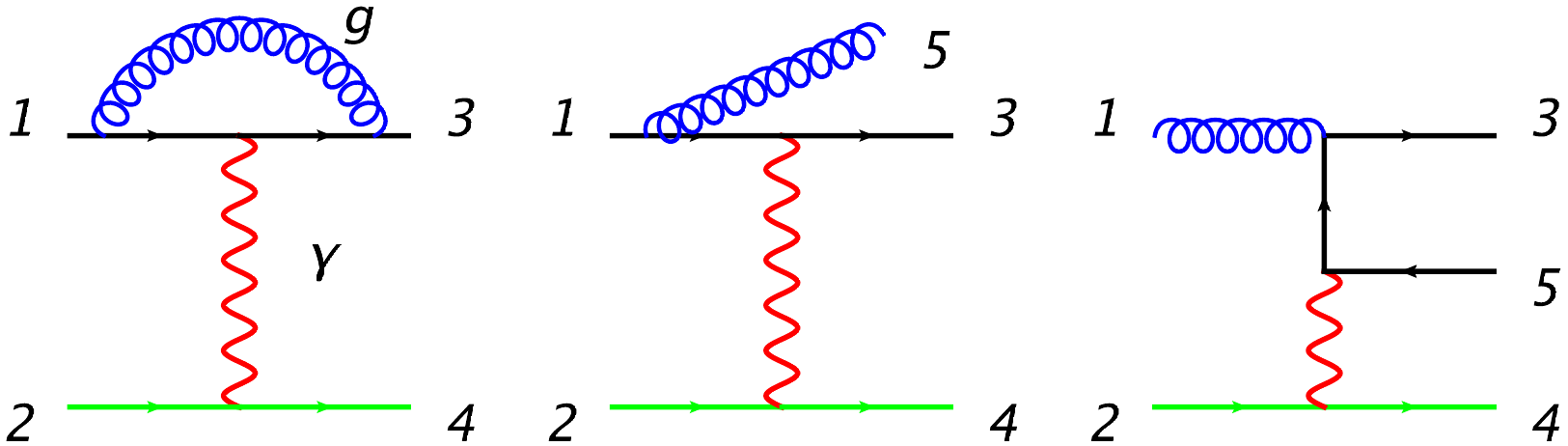}%
\caption{Representative Feynman diagrams contributing to the following perturbative QCD corrections at NLO: virtual corrections to the 
$q(p_1)+l(p_2) \to q(p_3)+l(p_4)$ process (left); real emission correction $q(p_1)+l(p_2) \to q(p_3)+l(p_4)+g(p_5)$ (middle); the process $g(p_1)+l(p_2) \to q_(p_3) +l(p_4)+\bar{q}(p_5)$ (right).  We have colored the photon line red, the lepton lines green, the gluon lines blue and the quark lines black.} \label{fig:NLOlepdiag}
\end{figure}

At the next-to-leading order level several new contributions first occur.  The quark-lepton scattering channel that appears at LO receives both virtual and real-emission corrections that are separately infrared divergent.  We use the antennae subtraction method~\cite{Kosower:1997zr} to regularize and cancel these divergences.  Initial-state collinear divergences are handled as usual by absorbing them into the PDFs via mass factorization.   A gluon-lepton scattering channel also contributes at this order.  The collinear divergences that appear in these contributions are removed by mass factorization.  Example Feynman diagrams for these processes are shown in Fig.~\ref{fig:NLOlepdiag}.

\begin{figure}[h]
\centering
\includegraphics[width=3.0in]{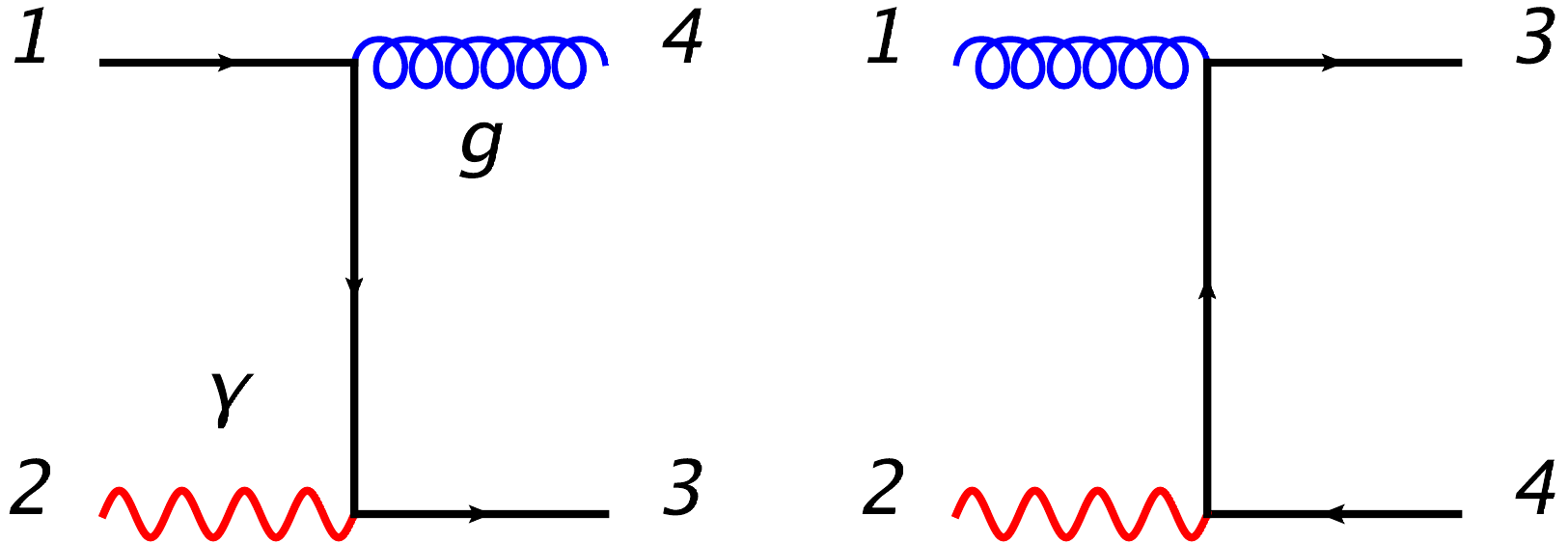}%
\caption{Representative Feynman diagrams contributing to the $q(p_1)+\gamma(p_2) \to q(p_3)+g(p_4)$ (left) and $g(p_1)+\gamma(p_2) \to q(p_3)+\bar{q}(p_4)$ scattering processes.} \label{fig:NLOphotdiag}
\end{figure}

The processes discussed above exhaust the possible NLO contributions when the final-state lepton is observed.  However,  for single-inclusive jet production a kinematic configuration is allowed where the $t$-channel photon is nearly on-shell, and the final-state lepton travels down the beam pipe. The transverse momentum of the leading jet is balanced by the additional jet present in these diagrams.  This kinematic configuration leads to a QED collinear divergence for vanishing lepton mass, since the photon can become exactly on-shell in this limit.  While it is regulated by the lepton mass, it is more convenient to obtain these corrections by introducing a photon distribution function in analogy with the usual parton distribution function.  The collinear divergences that appear in the matrix elements computed with vanishing lepton mass can be absorbed into this distribution function, which can be calculated in perturbation theory.  Representative diagrams for the photon-initiated processes are shown in Fig.~\ref{fig:NLOphotdiag}.
 
The full expression for the NLO hadronic cross section then takes the form
\beqa\label{eq:hxsecnlo}
\rd\sigma_{\NLO}&=&\int\frac{\rd\xi_1}{\xi_1}\frac{\rd\xi_2}{\xi_2}\bigg\{f_{g/H}^1f_{l/l}^2 \ds_{gl}^{(2,1)} + f_{g/H}^1f_{\gamma/l}^2\ds_{g\gamma}^{(1,1)} \phantom{\sum_q}\nn \\
&& +\sum_q\bigg[f_{q/H}^1f_{l/l}^2\ds_{ql}^{(2,1)}+f_{\bar{q}/H}^1f_{l/l}^2\ds_{\bar{q}l}^{(2,1)} \nn \\
&&+f_{q/H}^1f_{\gamma/l}^2\ds_{q\gamma}^{(1,1)}+f_{\bar{q}/H}^1f_{\gamma/l}^2\ds_{\bar{q}\gamma}^{(2,1)}\bigg]\bigg\},
\eeqa
where we have abbreviated $f_{i/j}^k=f_{i/j}(\xi_k)$.  The contributions $\ds_{gl}^{(2,1)}$, $\ds_{ql}^{(2,1)}$ and $\ds_{\bar{q}l}^{(2,1)}$ denote the usual DIS partonic channels computed to NLO in QCD with zero lepton mass.  The terms $\ds_{\bar{q}\gamma,\LO}$ and $\ds_{g\gamma,\LO}$ denote the new contributions arising when $Q^2 \approx 0$ and the virtual photon is nearly on-shell.  The photon distribution can be expressed as
\begin{equation}
f_{\gamma/l}(\xi) = \frac{\alpha}{2\pi} P_{\gamma l}(\xi) \left[ \text{ln}\left(\frac{\mu^2}{\xi^2 m_l^2} \right)-1\right] +{\cal O}(\alpha^2),
\end{equation}
where the splitting function is given by
\begin{equation}
\label{eq:gammalsplit}
P_{\gamma l}(\xi) = \frac{1+(1-\xi)^2}{\xi}.
\end{equation} 
This is the well-known Weizs\"{a}cker-Williams (WW) distribution for the photon inside of a lepton~\cite{WW}.  The appearance of the renormalization scale $\mu$ indicates that an $\overline{\text{MS}}$ subtraction of the QED collinear divergence is used in the calculation of the $gl$ and $ql$ scattering channels, and consequently in the derivation of the photon distribution function. 

\section{Calculation of the NNLO result} \label{sec:nnlo}

The calculation of the full ${\cal O}(\alpha^2\alpha_s^2)$ corrections involves several distinct contributions.  The quark-lepton and gluon-lepton scattering channels receive two-loop double virtual corrections, one-loop corrections to single real-emission diagrams, and double-real emission corrections.  These contributions necessitate the use of a full-fledged NNLO subtraction scheme.  We use the recently-developed $N$-jettiness subtraction scheme~\cite{Boughezal:2015dva,Gaunt:2015pea}.  Its application to this process is discussed here in detail.  In addition, the photon-initiated scattering channels receive virtual and single real-emission corrections.  The calculation of these terms follows the standard application of the antennae subtraction scheme at NLO. 

There is in addition a new effect that appears at the NNLO level.  The initial-state lepton can emit a photon which splits into a $q\bar{q}$ pair, all of which are collinear to the initial lepton direction.  In the limit of vanishing fermion masses there is a collinear singularity associated with this contribution.  This divergence appears in the quark-lepton, gluon-lepton, and photon-initiated scattering channels.  It can be absorbed into a distribution function that describes the quark distribution inside a lepton.  Treating the collinear singularity in this way leads to new scattering channels that first appear at NNLO: $q\bar{q} \to q\bar{q}$, $q\bar{q} \to q^{\prime}\bar{q}^{\prime}$, $q\bar{q} \to gg$, $qq^{\prime} \to qq^{\prime}$, and $qg \to qg$.  For our numerical predictions for these channels we need the quark distribution in a lepton.  We obtain this by solving the DGLAP equation, which to the order we are working can be written as
\beqa\label{eq.dglap}
&&\mu^2\frac{\partial f_{q/l}}{\partial \mu^2}(\xi,\mu^2)=e_q^2\frac{\alpha}{2\pi}\int_{\xi}^1\frac{\rd z}{z}P_{q\gamma}^{(0)}(z)\,f_{\gamma/l}\left(\frac{\xi}{z},\mu^2\right) \nn \\ &&+ e_q^2\left(\frac{\alpha}{2\pi}\right)^2\int_{\xi}^1\frac{\rd z}{z}P_{ql}^{(1)}(z)\,f_{l/l}\left(\frac{\xi}{z},\mu^2\right),
\eeqa
where the two needed splitting kernels are
\beqa
P_{q\gamma}^{(0)}(x)&=&x^2+(1-x)^2, \nn \\
P_{ql}^{(1)}(x) &=& -2+\frac{20}{9x}+6x-\frac{56x^2}{9}\nn \\ &&\hspace{-0.8cm}+\left(1+5x+\frac{8x^2}{3}\right) \log(x)-(1+x) \log^2(x).
\eeqa
This expression for the NLO splitting kernel can be obtained from upon replacement of the appropriate QCD couplings with electromagnetic ones.  To derive the full quark-in-lepton distribution we use as an initial condition $ f_{q/l}(\xi,m_l^2)=0$. Solving Eq.~(\ref{eq.dglap}) with this initial condition gives
\beqa
\label{eq:qinl}
&&f_{q/l}(\xi,\mu^2)=e_q^2\left(\frac{\alpha}{2\pi}\right)^2\bigg\{\bigg[\frac{1}{2}+\frac{2}{3\xi}-\frac{\xi}{2}-\frac{2\xi^2}{3} \nn \\ && +(1+\xi)\log\xi \bigg]  \log^2\left(\frac{\mu^2}{m_l^2}\right) + \bigg[-3-\frac{2}{\xi}+7\xi-2\xi^2+\nn \\ &&\bigg( -5 -\frac{8}{3\xi}+\xi+\frac{8\xi^2}{3}\bigg)\log(\xi)-3(1+x)\log^2(\xi)\bigg] \nn \\ && \times \log\left(\frac{\mu^2}{m_l^2}\right).
\eeqa

\begin{figure}[h]
\centering
\includegraphics[width=3.0in]{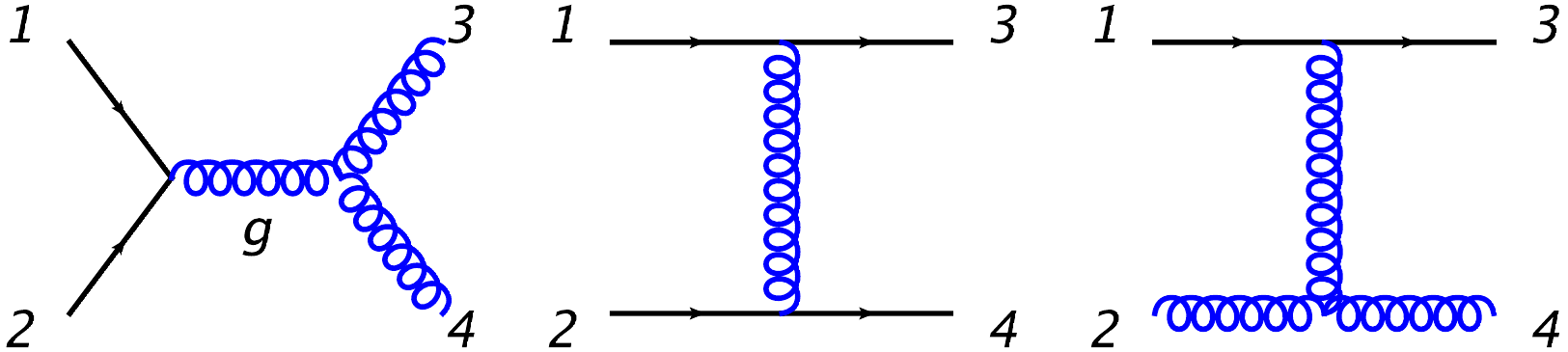}%
\caption{Representative Feynman diagrams contributing to the $q(p_1)+\bar{q}(p_2) \to g(p_3)+g(p_4)$ (left), $q(p_1)+q^{\prime}(p_2) \to q(p_3)+q^{\prime}(p_4)$ (middle), and $q(p_1)+g(p_2) \to q(p_3)+g(p_4)$ (right) scattering processes.} \label{fig:NNLO2jet}
\end{figure}

With this distribution function it is straightforward to obtain numerical predictions for these partonic channels.  Representative Feynman diagrams for these processes are shown in Fig.~\ref{fig:NNLO2jet}.  We can now write down the full result for the ${\cal O}(\alpha^2 \alpha_s^2)$ correction to the cross section:
\beqa
\label{eq:hxsecnnlo}
\rd\sigma_{\NNLO}&=&\int\frac{\rd\xi_1 \rd\xi_2}{\xi_1 \xi_2}\bigg\{f_{g/H}^1f_{l/l}^2\,\ds_{gl}^{(2,2)}
+f_{g/H}^1f_{\gamma/l}^2\,\ds_{g\gamma}^{(1,2)} \phantom{\sum_q}\nn \\
&+&\sum_q\bigg[f_{g/H}^1f_{q/l}^2\,\ds_{gq}^{(0,2)}+f_{g/H}^1f_{\bar{q}/l}^2\,\ds_{g\bar{q}}^{(0,2)}\bigg] \nn\\
&+&\sum_q\bigg[f_{q/H}^1f_{l/l}^2\,\ds_{ql}^{(2,2)}+f_{\bar{q}/H}^1f_{l/l}^2\,\ds_{\bar{q}l}^{(2,2)} \nn \\
&+&f_{q/H}^1f_{\gamma/l}^2\,\ds_{q\gamma}^{(1,2)}+f_{\bar{q}/H}^1f_{\gamma/l}^2\,\ds_{\bar{q}\gamma}^{(1,2)} \phantom{\sum_q\bigg[} \nn \\
&+&f_{q/H}^1f_{\bar{q}/l}^2\,\ds_{q\bar{q}}^{(0,2)}+f_{\bar{q}/H}^1f_{q/l}^2\,\ds_{\bar{q}q}^{(0,2)} \bigg] \phantom{\sum_q} \nn \\
&+&\sum_{q,q'}\bigg[f_{q/H}^1f_{q'/l}^2\,\ds_{qq'}^{(0,2)}+f_{\bar{q}/H}^1f_{\bar{q}'/l}^2\,\ds_{\bar{q}\bar{q}'}^{(0,2)}  \nn \\
&+&f_{q/H}^1f_{\bar{q}'/l}^2\,\ds_{q\bar{q}'}^{(0,2)}+f_{\bar{q}/H}^1f_{q'/l}^2\,\ds_{\bar{q}q'}^{(0,2)}\bigg]\bigg\}. \phantom{\sum_q}
\eeqa

\begin{figure}[h]
\centering
\includegraphics[width=3.0in]{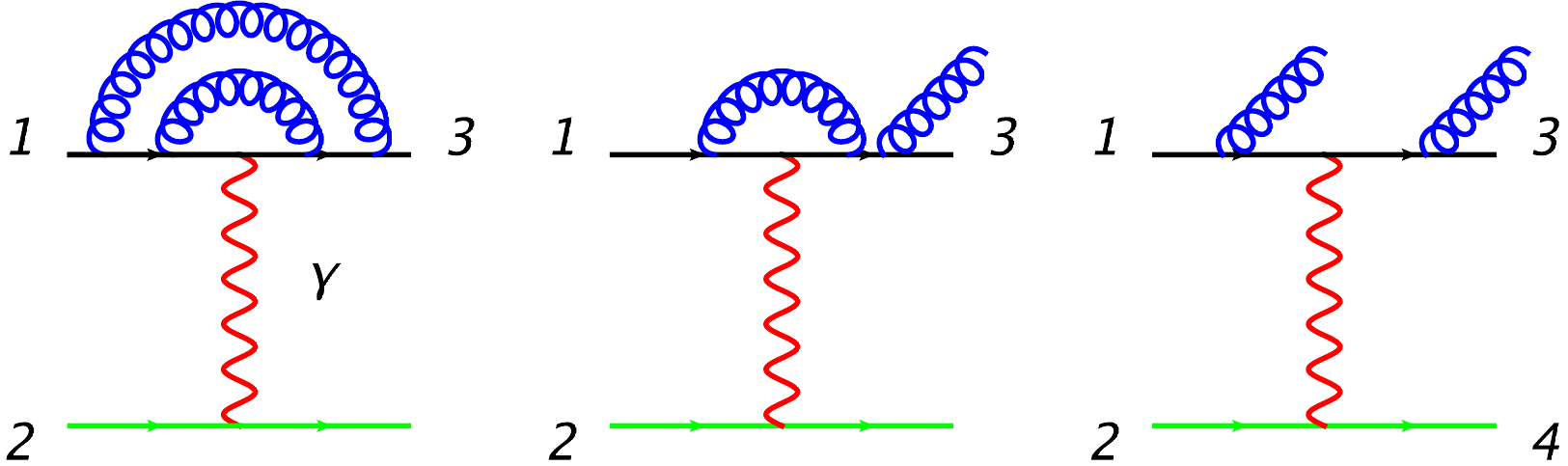}%
\caption{Representative Feynman diagrams contributing to the quark-lepton scattering channel at NNLO.} \label{fig:NNLOql}
\end{figure}

The most difficult contribution is the quark-lepton scattering channel.  It receives contributions from two-loop virtual corrections (double-virtual), one-loop corrections to single real emission terms (real-virtual), and double-real emission corrections.  Sample Feynman diagrams for these corrections are shown in Fig.~\ref{fig:NNLOql}.  These are separately infrared divergent, and require a full NNLO subtraction scheme to combine.  We apply the $N$-jettiness subtraction scheme~\cite{Boughezal:2015dva,Gaunt:2015pea}.  The starting point of this method is the $N$-jettiness event shape variable~\cite{Stewart:2010tn}, defined in the one-jet case of current interest as
\begin{equation}
\label{eq:taudef}
\tauone = \frac{2}{Q^2} \sum_i \text{min}\left\{p_B \cdot q_i, p_J \cdot q_i \right\},
\end{equation}
with $Q^2 = -(p_2-p_4)^2$.  Here, $p_B$ and $p_J$ are light-like four-vectors along the initial-state hadronic beam and final-state jet directions, respectively.\footnote{This choice of $\tauone$ corresponds to $\tau_1^a$ in Ref.~\cite{Kang:2013nha}.  We note that this definition is dimensionless, unlike the choice in previous applications of $N$-jettiness subtraction~\cite{Boughezal:2015dva}.}  The $q_i$ denote the four-momenta of all final-state partons.   Values of $\tauone$ near zero indicate a final state containing a single narrow energy deposition, while larger values denote a final state containing two or more well-separated energy depositions.  A measurement of $\tauone$ is itself of phenomenological interest.  It has been proposed as a probe of nuclear properties in electron-ion collisions~\cite{Kang:2012zr,Kang:2013wca}, and has also been suggested as a way to precisely determine the strong coupling constant~\cite{Kang:2013nha}.

We will use $\tauone$ to establish the complete ${\cal O}(\alpha^2\alpha_s^2)$ calculation of  the quark-lepton  scattering channel.  Our ability to do so relies on two key observations, as first discussed in Ref.~\cite{Boughezal:2015dva} for a general $N$-jet process.

\begin{itemize}

\item Restricting $\tauone >0$ removes all doubly-unresolved limits of the quark-lepton matrix elements, for example when the two additional partons that appear in the double-real emission corrections are soft or collinear to the beam or the final-state jet.  This can be seen from Eq.~(\ref{eq:taudef}); if $\tauone >0$ then at least one $q_i$ must be resolved.  Since all doubly-unresolved limits are removed, the ${\cal O}(\alpha^2\alpha_s^2)$ correction in this phase space region can be obtained from an NLO calculation of two-jet production in electron-nucleon collisions.  

\item When $\tauone$ is smaller than any other hard scale in the problem, it can be resummed to all orders in perturbation theory~\cite{Stewart:2009yx,Stewart:2010pd}.  Expansion of this resummation formula to ${\cal O}(\alpha^2\alpha_s^2)$ gives the NNLO correction to the quark-lepton scattering channel for small $\tauone$.

\end{itemize}

The path to a full NNLO calculation is now clear.   We partition the phase space for the real-virtual and double-real corrections
into regions above and below a cutoff on $\tauone$, which we label $\tauonecut$:
\begin{equation}
\label{eq:partition}
\begin{split}
\rd\sigma_{ql}^{(2,2)} &= \int {\rm d}\Phi_{\text{VV}} \, |{\cal M}_{\text{VV}}|^2 +\int {\rm d}\Phi_{\text{RV}} \, |{\cal M}_{\text{RV}}|^2 \, \theta_1^{<} \\
&+\int {\rm d}\Phi_{\text{RR}} \, |{\cal M}_{\text{RR}}|^2 \, \theta_1^{<}+\int {\rm d}\Phi_{\text{RV}} \, |{\cal M}_{\text{RV}}|^2 \, \theta_1^{>} \\
&+\int {\rm d}\Phi_{\text{RR}} \, |{\cal M}_{\text{RR}}|^2 \, \theta_1^{>} \\
 \equiv & \rd\sigma_{ql}^{(2,2)}(\tauone < \tauonecut)+\rd\sigma_{ql}^{(2,2)}(\tauone > \tauonecut)
\end{split}
\end{equation}
We have abbreviated $\theta_1^{<} = \theta(\tauonecut-\tauone)$ and $\theta_1^{>} = \theta(\tauone-\tauonecut)$.  The first three terms in this expression all have $\tauone<\tauonecut$, and have been collectively denoted as $\rd\sigma_{ql}^{(2,2)}(\tauone < \tauonecut)$.  The remaining two terms have $\tauone>\tauonecut$, and have been collectively denoted as $\rd\sigma_{ql}^{(2,2)}(\tauone > \tauonecut)$.  The double-virtual corrections necessarily have $\tauone=0$.  We obtain $\rd\sigma_{ql}^{(2,2)}(\tauone > \tauonecut)$ from a NLO calculation of two-jet production.  This is possible since no genuine double-unresolved limit occurs in this phase-space region.  We discuss the calculation of  $\rd\sigma_{ql}^{(2,2)}(\tauone < \tauonecut)$ using the all-orders resummation of this process in the following sub-section.  We note that only the quark-lepton and gluon-lepton scattering channels have support for $\tauone=0$.  For the other processes there are two final-state jets with non-zero transverse momentum.  Such configurations necessarily have $\tauone >0$, and therefore only receive contributions from the 
above-the-cut phase-space region.  We only need these contributions to at most NLO in QCD perturbation theory, which can be obtained via standard techniques.  We use antennae subtraction.  The only non-standard aspect of this NLO calculation is the appearance of triple-collinear QED limits associated with the emission of a photon and a $q\bar{q}$ pair which require the use of integrated antennae found in Refs.~\cite{Daleo:2009yj,Boughezal:2010mc,GehrmannDeRidder:2012ja}.  A powerful aspect of the $N$-jettiness subtraction method is its ability to upgrade existing NLO calculations to NNLO precision.   Previous applications of $N$-jettiness subtraction~\cite{Boughezal:2015dva,Boughezal:2015aha,Boughezal:2015ded} have used the NLO dipole subtraction technique~\cite{Catani:1996vz} to facilitate the calculation of the above-cut phase-space region.  This work demonstrates that it can also be used in conjunction with the NLO antennae subtraction scheme.


\subsubsection{Below $\tauonecut$}

An all-orders resummation of the $\tauone$ event-shape variable in the DIS process for the limit $\tauone \ll 1$ was constructed in Refs.~\cite{Kang:2013wca,Kang:2013nha}:
\begin{eqnarray}
\label{eq:SCETfac}
\frac{\rd \sigma}{\rd \tauone} &=& \int \rd  \Phi_2(p_3,p_4;p_1,p_2) \int \rd t_J \rd t_B \rd k_S  \nn \\ &\times & \delta\left( \tauone-\frac{t_J}{Q^2}-\frac{t_B}{Q^2} -\frac{k_S}{Q} \right)  \nonumber \\ &\times & \sum_q J_q(t_J,\mu) \,S(k_S,\mu)
 H_q(\Phi_2,\mu) B_q(t_B,x,\mu)+ \ldots \nn \\
\end{eqnarray}  
We have allowed the index $q$ to run over both quarks and anti-quarks.  $x$ denotes the usual Bjorken scaling variable for DIS:
\begin{equation}
x = \frac{Q^2}{2 P\cdot (p_2-p_4)},
\end{equation}
where $P$ is the initial-state nucleon four-momentum.  $\Phi_2$ denotes the Born phase space, which consists of a quark and a lepton.  The derivation of this result relies heavily on the machinery of soft-collinear effective theory (SCET)~\cite{Bauer:2000ew}.  A summary of the SCET functions that appear in this expression and what they describe is given below.
\begin{itemize}

\item $H$ is the hard function that encodes the effect of hard virtual corrections.  At leading order in its $\alpha_s$ expansion it reduces to the leading-order partonic cross section.  At higher orders it also contains the finite contributions of the pure virtual corrections, renormalized using the $\overline{\text{MS}}$ scheme.  It depends only on the Born-level kinematics and on the scale choice.

\item $J_q$, the quark jet function, describes the effect of radiation collinear to the final-state jet (which for this process is initiated by a quark at LO).  It depends on $t_J$, the contribution of final-state collinear radiation to $\tauone$.  It possesses a perturbative expansion in $\alpha_s$.

\item $S$ is the soft function that encodes the contributions of soft radiation.   It depends on $k_S$, the contribution of soft radiation to $\tauone$, and has a perturbative expansion in $\alpha_s$.

\item $B$ is the beam function that contains the effects of initial-state collinear radiation.  It depends on $t_B$, the contribution of initial-state collinear radiation to $\tauone$.  The beam function is non-perturbative; however, up to corrections suppressed by $\Lambda^2_{\text{QCD}}/t_B$, it can be written as a convolution of perturbative matching coefficients and the usual PDFs:
\begin{equation}
B_q(t_B,x,\mu) = \sum_i \int_x^1 \frac{\rd \xi}{\xi} {\cal I}_{qi} (t_B,x/\xi,\mu) f_{i/H}(\xi),
\end{equation}
where we have suppressed the scale dependence of the PDF, and $i$ runs over all partons.

\end{itemize}
The delta function appearing in Eq.~(\ref{eq:SCETfac}) combines the contribution of each type of radiation to produce the measured value of $\tauone$.  The ellipsis denotes power corrections that are small as long as we restrict ourselves to the phase-space region $\tauone \ll 1$. 

The hard, jet, and soft functions as well as the beam-function matching coefficients all have perturbative expansions in $\alpha_s$ that can be obtained from the literature~\cite{Zijlstra:1992qd,Becher:2006qw,Gaunt:2014xga,Boughezal:2015eha}.  Upon expansion to ${\cal  O}(\alpha_s^2)$ and integration over the region $\tauone < \tauonecut$, Eq.~(\ref{eq:SCETfac}) will give exactly the cross section $\rd\sigma_{ql}^{(2,2)}(\tauone < \tauonecut)$ that we require.  We can match the beam function onto the PDFs and rewrite the cross section below the cut as
\begin{eqnarray}
\rd\sigma &=& \int \frac{\rd \xi_1}{\xi_1} \frac{\rd \xi_2}{\xi_2} \sum_{q,i}  f_{i/H}(\xi_1) f_{l/l}(\xi_2)  \int \rd  \Phi_2(p_3,p_4;p_1,p_2) \nn \\ && \int_0^{\tauonecut} \rd \tauone \int \rd t_J \rd t_B \rd k_S \delta\left( \tauone-\frac{t_J}{Q^2}-\frac{t_B}{Q^2} -\frac{k_S}{Q} \right)  \nonumber \\ &\times &  J_q(t_J,\mu) \,S(k_S,\mu)
 H_q(\Phi_2,\mu) \, {\cal I}_{qi} (t_B,x/\xi,\mu) \nonumber \\
&\equiv& \sum_{q,i} \int \rd\Phi^{i}_{\text{Born}} \, \left[ J_q \otimes S \otimes H_q \otimes {\cal I}_{qi} \right].
\end{eqnarray}
We have introduced the schematic notation $J_q \otimes S \otimes H_q \otimes {\cal I}_{qi}$ for the integrations of the SCET functions over $\tauone$, $t_J$, $t_B$, and $k_S$; $\rd\Phi^{i}_{\text{Born}}$ represents all other terms for the given index $i$: the parton distribution functions, the integral over the Born phase space, and any measurement function acting on the Born variables.  We denote the expansion of these functions in $\alpha_s$ as
\begin{equation}
{\cal X} = {\cal X}^{(0)}+{\cal X}^{(1)}+{\cal X}^{(2)}+\ldots,
\end{equation}
where the superscript denotes the power of $\alpha_s$ appearing in each term.  With this notation, we need the following contributions to obtain the ${\cal O}(\alpha_s^2)$ correction to the cross section below $\tauonecut$: 
\begin{eqnarray}
\label{eq:belowcutexp}
\rd\sigma_{\text{NNLO}} &=& \sum_{q,i} \int \rd\Phi^i_{\text{Born}} \, \bigg\{J_q^{(2)} \otimes S^{(0)} \otimes H_q^{(0)} \otimes {\cal I}_{qi}^{(0)} \nn \\
	&+&J_q^{(0)} \otimes S^{(2)} \otimes H_q^{(0)} \otimes {\cal I}_{qi}^{(0)} + J_q^{(0)} \otimes S^{(0)} \otimes H_q^{(2)} \nn \\ &\otimes & 
	{\cal I}_{qi}^{(0)}+J_q^{(0)} \otimes S^{(0)} \otimes H_q^{(0)} \otimes {\cal I}_{qi}^{(2)} \nonumber \\ 
	&+&J_q^{(1)} \otimes S^{(1)} \otimes H_q^{(0)} \otimes {\cal I}_{qi}^{(0)}+J_q^{(1)} \otimes S^{(0)} \otimes H_q^{(1)} \nn \\ &\otimes & {\cal I}_{qi}^{(0)}
	+J_q^{(1)} \otimes S^{(0)} \otimes H_q^{(0)} \otimes {\cal I}_{qi}^{(1)}+J_q^{(0)} \otimes  S^{(1)}  \nn \\ & \otimes &H_q^{(1)} \otimes  {\cal I}_{qi}^{(0)} 
	+ J_q^{(0)} \otimes S^{(1)} \otimes H_q^{(0)} \otimes {\cal I}_{qi}^{(1)}\nn \\ &+& J_q^{(0)} \otimes S^{(0)} \otimes H_q^{(1)} \otimes {\cal I}_{qi}^{(1)} \bigg\}.
\end{eqnarray}

To simplify this expression, we first note that the hard function has no dependence on the hadronic variables $\tauone$, $t_J$, $t_B$, and $k_S$.  It depends only on the Born phase space and is a multiplicative factor for the hadronic integrations.  Next, we note that the leading-order expressions for the SCET functions are proportional to delta functions in their respective hadronic variable:
\begin{equation}
{\cal X}^{(0)} \propto \delta(t_{\cal X})
\end{equation} 
for ${\cal X}=J_q, S$, or ${\cal I}_{qi}$. This simplifies the integrals involving an ${\cal X}^{(2)}$.  Using the $J_q^{(2)}$ term in Eq.~(\ref{eq:belowcutexp}) as an example, we have
\begin{equation}
J_q^{(2)} \otimes S^{(0)} \otimes H_q^{(0)} \otimes {\cal I}_{qi}^{(0)} = Q^2 H_q^{(0)} \delta_{qi}\, \int_0^{\tauonecut} \hspace{-0.5cm}\rd \tauone J_q^{(2)} (\tauone Q^2,\mu),
\end{equation}
where the $\delta_{qi}$ comes from the ${\cal I}_{qi}^{(0)}$ term.  Using the fact that the jet function can be written in the form 
\begin{equation}
J_q^{(2)} (t_J,\mu) = a_{-1} \delta(t_J)+\sum_{n=0}^3 a_n \, \frac{1}{\mu^2} \left[  \frac{\mu^2 \,\text{ln}^n(t_J/\mu^2)}{t_J}\right]_+,
\end{equation}
where the $a_i$ denote coefficients that can be found in the literature~\cite{Becher:2006qw}, we can immediately derive 
\begin{equation}
\begin{split}
J_q^{(2)} \otimes S^{(0)} \otimes H_q^{(0)} & \otimes {\cal I}_{qi}^{(0)} = H_q^{(0)} \,\delta_{qi} \left\{ a_{-1} \right. \\ & \left. +\sum_{n=0}^3 \frac{1}{n+1} a_{n+1} \,\text{ln}^{n+1} \left( \frac{\tauonecut Q^2}{\mu^2}\right)\right\}.
\end{split}
\end{equation} 
We have used the standard definition of the plus distribution of a function:
\begin{equation}
\int_0^1 \rd x \, \left[f(x)\right]_+ \, g(x) = \int_0^1 \rd x \, f(x) \left[g(x)-g(0) \right].
\end{equation}
This analysis shows how to analytically calculate any term containing one of the NNLO SCET functions.  

It remains only to calculate contributions containing two NLO SCET functions.  We focus on $J_q^{(1)} \otimes S^{(1)} \otimes H_q^{(0)} \otimes {\cal I}_{qi}^{(0)} $ as an example.  Using the LO expression for the beam function we can immediately derive the equation
\begin{equation}
\begin{split}
J_q^{(1)} \otimes S^{(1)} \otimes H_q^{(0)} &\otimes {\cal I}_{qi}^{(0)} = Q \, H_q^{(0)} \,\delta_{qi}\, \int_0^{\tauonecut} \hspace{-0.5cm}\rd \tauone \int_0^{\tauone Q} dt_J \\ & \times
	J_q^{(1)} (t_J,\mu) \, S^{(1)}\left(Q\tauone-\frac{t_J}{Q} \right).
\end{split}
\end{equation}
The NLO results for the jet and soft functions can be written as
\begin{eqnarray}
J_q^{(1)} (t_J,\mu) &=& b_{-1} \delta(t_J)+\sum_{n=0}^1 b_n \, \frac{1}{\mu^2} \left[  \frac{\mu^2 \,\text{ln}^n(t_J/\mu^2)}{t_J}\right]_+, \nonumber \\
S^{(1)} (k_S,\mu) &=& c_{-1} \delta(k_S)+\sum_{n=0}^1 c_n \, \frac{1}{\mu} \left[  \frac{\mu \,\text{ln}^n(k_S/\mu)}{k_S}\right]_+ .
\end{eqnarray} 
Using these expressions it is straightforward to derive the result
\begin{eqnarray}
J_q^{(1)} &\otimes& S^{(1)} \otimes H_q^{(0)} \otimes {\cal I}_{qi}^{(0)} = H_q^{(0)} \,\delta_{qi} \bigg\{ b_{-1} c_{-1} 
\nn \\ &+& c_{-1} \, \sum_{n=0}^1 \frac{1}{n+1} \,b_{n+1}\, L_J^{n+1} 
+ b_{-1} \,  \sum_{n=0}^1 \frac{1}{n+1} \,c_{n+1}\, L_S^{n+1}
\nn \\ &+&\left. \sum_{n,m=0}^1 b_m c_n \left[ L_J^{\alpha} \,L_S^{\beta} \,\frac{\Gamma(\alpha)\Gamma(\beta)}{\Gamma(1+\alpha+\beta)}\right] \bigg|_{\alpha^m,\beta^n}\right\}.
\end{eqnarray}
The vertical bar in the last term indicates that we should take the $\alpha^m \beta^n$ coefficient of the series expansion of the bracketed term.  We have introduced the abbreviations
\begin{equation}
L_J = \text{ln} \left( \frac{\tauonecut Q^2}{\mu^2}\right), \;\;\; L_S = \text{ln} \left( \frac{\tauonecut Q}{\mu}\right).
\end{equation}
Using these results it is straightforward to analytically compute all of the necessary hadronic integrals in Eq.~(\ref{eq:belowcutexp}).  The remaining integrals are over the Born phase space and parton distribution functions, and are simple to perform numerically.   This completes the calculation of the $\tauone < \tauonecut$ phase space region.  We note that the cross section below $\tauonecut$ will contain terms of the form $\text{ln}^n (\tauonecut)$, where $n$ ranges from 0 to 4 at NNLO.  An important check of this framework is the cancellation of these terms against the identical logarithms that appear for $\tauone > \tauonecut$.  We must also choose $\tauonecut$ small enough to avoid the power corrections in Eq.~(\ref{eq:SCETfac}) that go as $\tauonecut/Q$. Both of these issues will be addressed in our later section on numerical results.

\subsubsection{Above $\tauonecut$}

We now briefly outline the computation of the quark-lepton channel in the phase space region $\tauone > \tauonecut$.    As discussed above this can be obtained from a NLO calculation of the DIS process with an additional jet.  In addition to the virtual corrections to the $q\,l \to q\,l\,g$ partonic process, there are numerous radiation processes that also contribute: $q\,l \to q\,l\,g\,g$,  $q\,l \to q\,l\,q'\,\bar{q}'$ and $q\,l \to q\,l\,q\,\bar{q}$.  In addition to the usual ultraviolet renormalization of the strong coupling constant, the real and virtual corrections are separately infrared divergent.  Remaining divergences after introducing an NLO subtraction scheme are associated with initial-state collinear singularities, and are handled via 
mass factorization.

\section{Validation and numerical results} \label{sec:numerics}

We have implemented the NNLO cross section of Eq.~(\ref{eq:hxsecnnlo}), as well as the LO and NLO results of Eqs.~(\ref{eq:sigLO}) and~(\ref{eq:hxsecnlo}) in a numerical code {\tt DISTRESS} that allows for arbitrary cuts to be imposed on the final-state lepton and jets.  We describe below the checks we have performed on our calculation.

The antennae subtraction method provides an analytic cancellation of the $1/\epsilon$ poles that appear in an NLO calculation.  We are therefore able to check this cancellation of poles for all components computed in this way.  This includes the entire $\sigma_{\NLO}$, as well as the following contributions to the NNLO hadronic cross section: $\ds_{g\gamma}^{(1,2)}$ and $\ds_{q\gamma}^{(1,2)}$.  The various contributions $\ds_{ij}^{(0,2)}$ that occur in $\sigma_{\NNLO}$ are finite, and simple to obtain.  Our NLO results for the transverse momentum distribution and pseudorapidity distribution of the leading jet are compared against the plots of Ref.~\cite{Hinderer:2015hra}.  We find good agreement with these results.

This leaves only the validation of the $\ds_{gl}^{(2,2)}$ and $\ds_{ql}^{(2,2)}$ contributions, which both utilize the $N$-jettiness subtraction technique.  There are two primary checks that these pieces must satisfy.  First, they must be independent of the parameter $\tauncut$.  This checks the implementation of the beam, jet and soft functions, which have logarithmic dependence on this parameter.  It also determines the range of $\tauncut$ for which the power corrections denoted by the ellipsis in Eq.~(\ref{eq:SCETfac}) are negligibly small.  Second, upon integration over the final-state hadronic phase space we must reproduce the NNLO structure functions first determined in Ref.~\cite{Zijlstra:1992qd}.  This is an extremely powerful check on our calculation, which essentially cannot be passed if any contribution is implemented incorrectly.

\begin{figure}[h]
\centering
\includegraphics[width=3.6in]{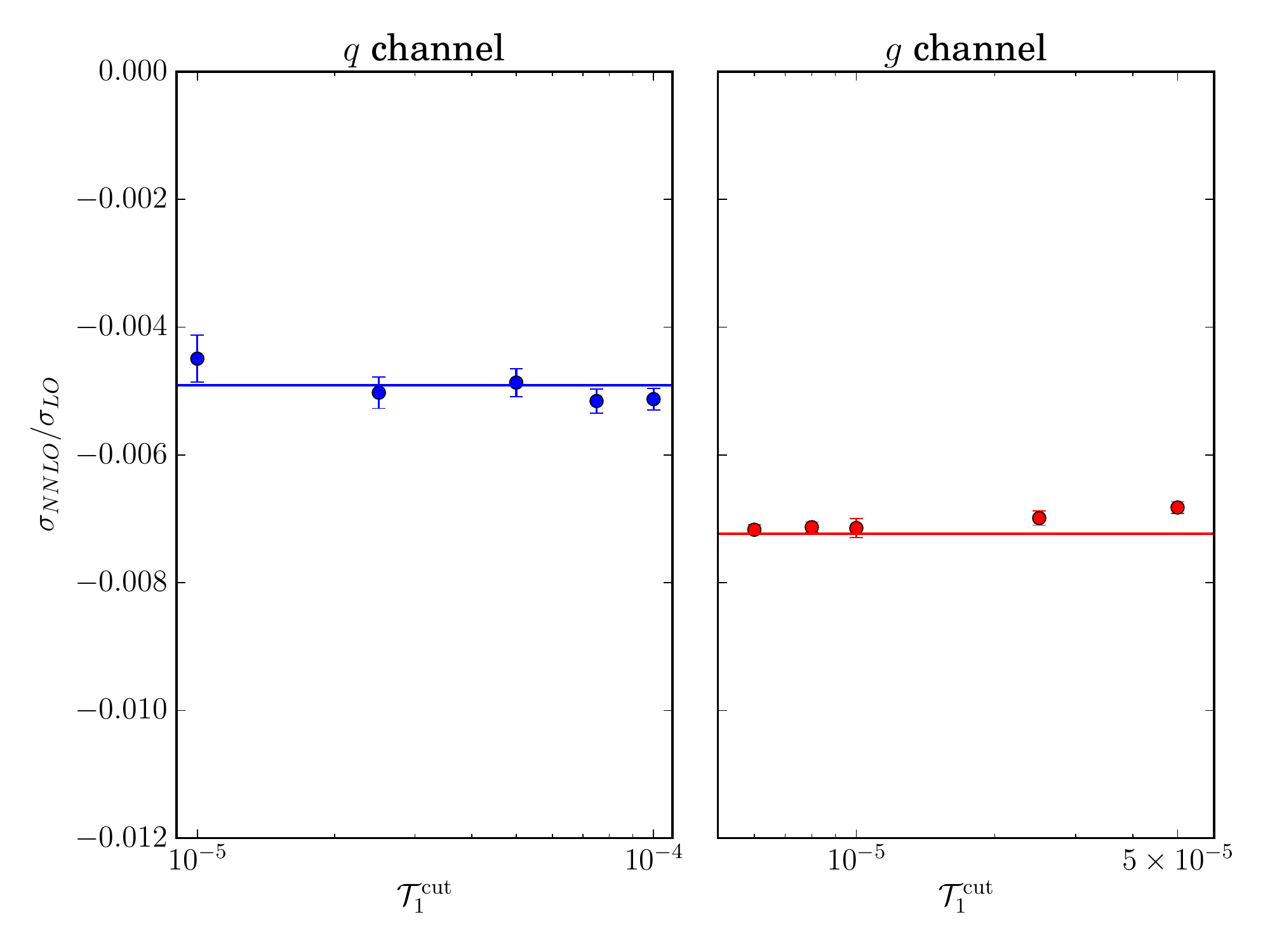}%
\caption{Plot of the NNLO corrections normalized to the LO cross section for the quark-lepton and gluon-lepton scattering channels as a function of $\tauonecut$.  The points denote values obtained from $N$-jettiness subtraction, with the vertical error bars denoting the numerical errors, while the solid lines indicate the inclusive structure function result. } \label{fig:valid}
\end{figure}

We show in Fig.~\ref{fig:valid} the results of these checks for the $ql$ and $gl$ scattering channels.  We have set the total center-of-mass energy of the lepton-proton collision to 100 GeV.  For the purpose of this validation check only we have imposed the phase-space cut $Q^2>100$ GeV$^2$, and have integrated inclusively over the hadronic phase space.  We have equated the renormalization and factorization scales to the common value $\mu=Q$, and have used the CT14 NNLO parton distribution functions~\cite{Dulat:2015mca}.  $\tauonecut$ has been varied from $5 \times 10^{-6}$ to $1 \times 10^{-4}$, and the ratio of the NNLO correction to the LO result for the cross section is shown.  The solid lines show the prediction of the inclusive structure function.  We first note that correction is extremely small, less than 1\% of the leading-order result.  Nevertheless we have excellent numerical control over the NNLO coefficient, as indicated by the vertical error bars.  Our numerical error on the NNLO coefficient is at the percent-level, sufficient for 0.01\% precision on the total cross section.  The $N$-jettiness prediction for the $ql$ scattering channel is independent of $\tauonecut$ over the studied range, while the $gl$ scattering channel is independent of $\tauonecut$ for $\tauonecut < 10^{-5}$. Both channels are in excellent agreement with the structure-function result.  We have also checked bin-by-bin that the transverse momentum and pseudorapidity distributions of the jet have no dependence on $\tauonecut$.  

\begin{figure}[h]
\centering
\includegraphics[width=3.65in]{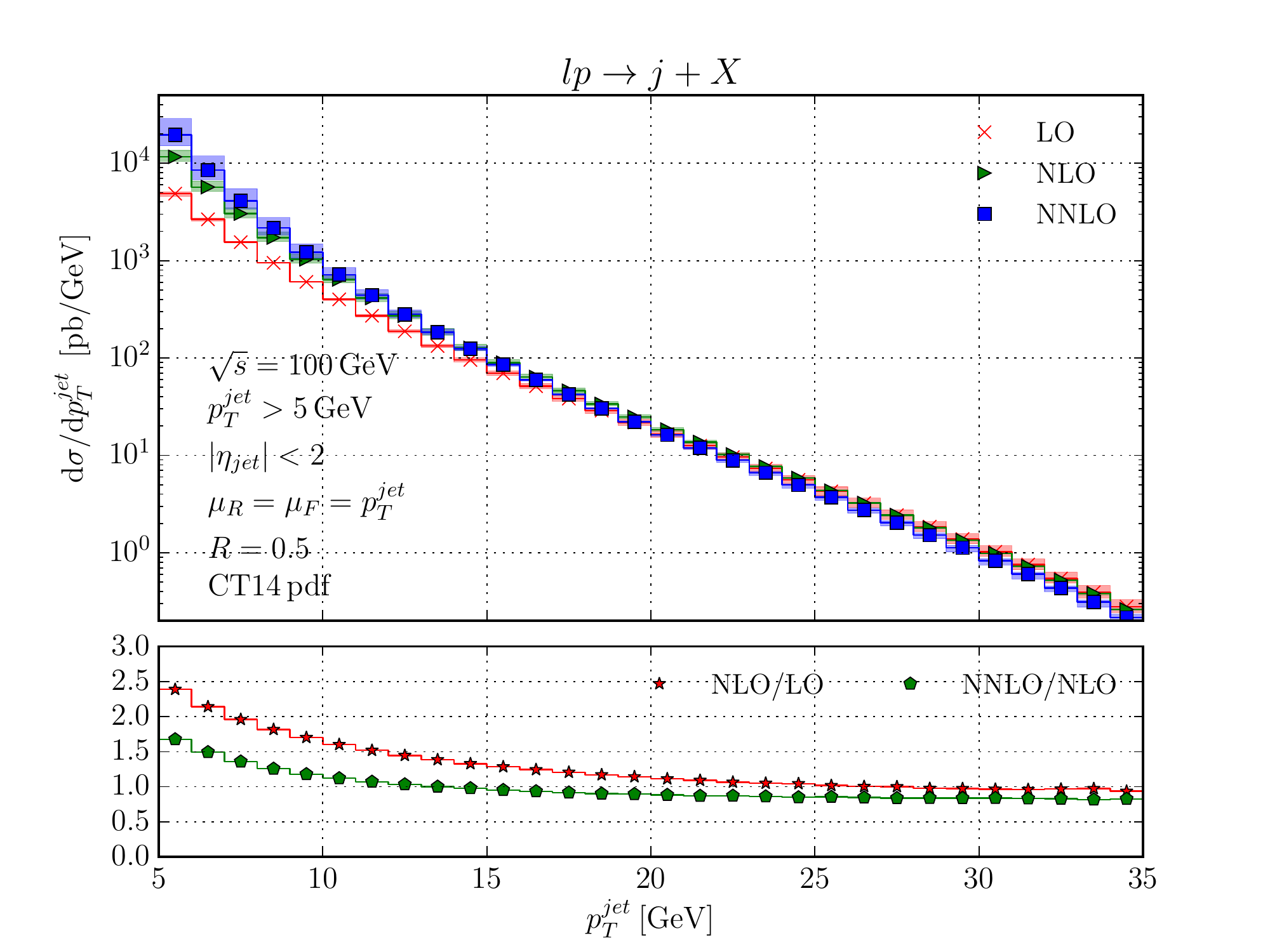}%
\caption{Plot of the inclusive-jet transverse momentum distribution at LO, NLO and NNLO in QCD perturbation theory.  The upper panel shows the distributions with scale uncertainties, while the lower panel shows the $K$-factors for the central scale choice.} \label{fig:ptj}
\end{figure}

Having established the validity of our calculation we present phenomenological results for proposed EIC run parameters.  We set the collider energy to $\sqrt{s}=100$ GeV and study the inclusive-jet transverse momentum and pseudorapidity distributions in the range $p_T^{jet}>5$ GeV and $|\eta_{jet}|<2$.  We use the CT14 parton distribution set~\cite{Dulat:2015mca} extracted to NNLO in QCD perturbation theory.  We reconstruct jets using the anti-$k_t$ algorithm~\cite{Cacciari:2008gp} with radius parameter $R=0.5$.  Our central scale choice for both the renormalization and factorization scales is $\mu=p_T^{jet}$.  To estimate the theoretical errors from missing higher-order corrections we vary the scale around its central value by a factor of two.  The transverse momentum distributions at LO, NLO and NNLO are shown in Fig.~\ref{fig:ptj}.  The $K$-factors, defined as the ratios of higher order over lower order cross sections, are shown in the lower panel of this figure.  The NNLO corrections are small in the region $p_T^{jet}>10$ GeV , changing the NLO result by no more than 10\% over the studied $p_T^{jet}$ region.  The shift of the NLO cross section is slightly positive in the low transverse momentum region, and become less than unity at high-$p_T^{jet}$.  Both the NNLO corrections and the scale dependence grow large at low-$p_T^{jet}$.  The large scale dependence arises primarily from the partonic channels $qq$ and $gq$.  These channels are effectively treated at leading order in our calculation, since they first appear at ${\cal O}(\alpha^2\alpha_s^2)$, and they are evaluated at the low scale $\mu=p_T^{jet}/2$ in our estimate of the theoretical uncertainty.  It is therefore not surprising that their uncertainty dominates at low-$p_T^{jet}$  These channels do not contribute at NLO, and consequently the NLO scale uncertainty is smaller.  This is an example of the potential pitfalls in using the scale uncertainty as an estimate of the theoretical uncertainty.  Only an explicit calculation can reveal qualitatively new effects that occur at higher orders in perturbation theory. 

\begin{figure}[h]
\centering
\includegraphics[width=3.65in]{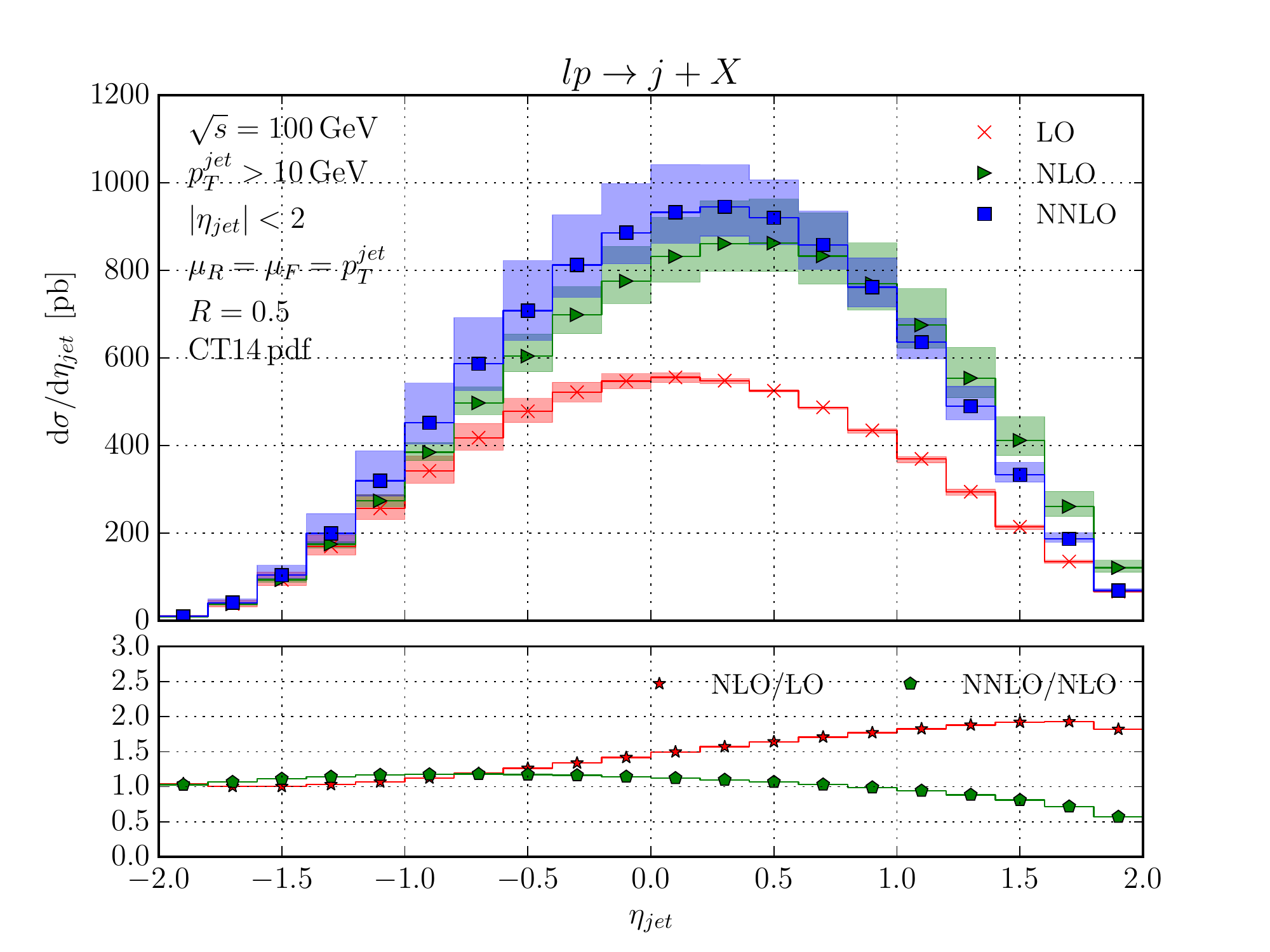}%
\caption{Plot of the inclusive-jet pseudorapidity distribution at LO, NLO and NNLO in QCD perturbation theory.  The upper panel shows the distributions with scale uncertainties, while the lower panel shows the $K$-factors for the central scale choice.} \label{fig:etaj}
\end{figure}

\begin{figure}[h]
\centering
\includegraphics[width=3.65in]{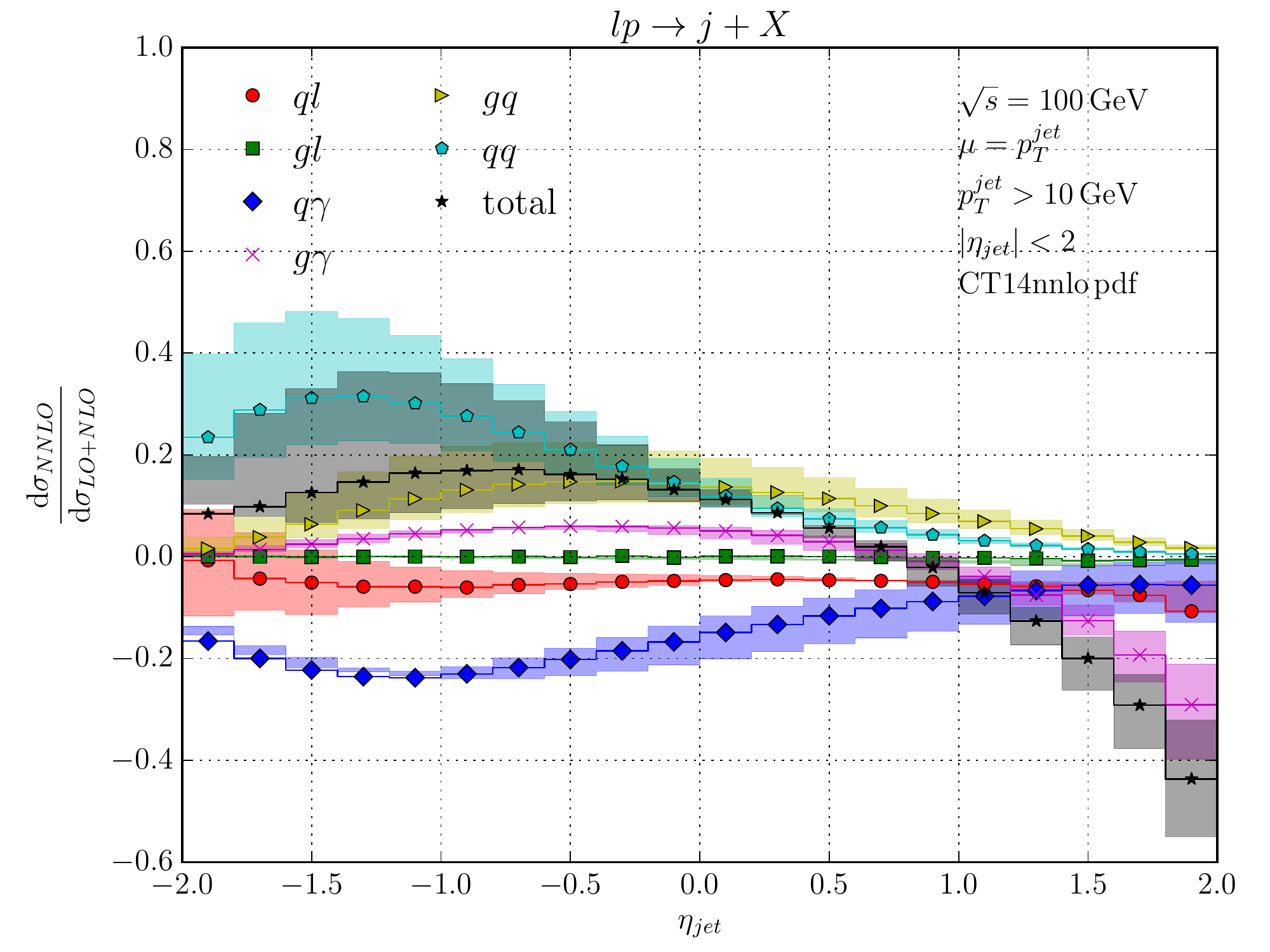}%
\caption{Breakdown of the NNLO correction to the $\eta_{jet}$ distribution into its constituent partonic channels, as a ratio to the full NLO cross section in the bin under consideration.  Also shown is the total result obtained by summing all channels. The bands indicate the scale variation. } \label{fig:eta_breakdown_muv}
\end{figure}

We next show the pseudorapidity distribution in Fig.~\ref{fig:etaj}, with the restriction $p_T^{jet}>10$ GeV.  There are a few surprising aspects present in the NNLO corrections.  First, the scale dependence at NNLO in the region $\eta_{jet}<0$ is larger than the corresponding NLO scale variation.  Although the corrections are near unity over most of the studied pseudorapidity range, they become sizable near $\eta_{jet} \approx 2$, reducing the NLO rate by nearly 50\%.  To determine the origin of these effects we show in Fig.~\ref{fig:eta_breakdown_muv} the breakdown of the NNLO correction into its separate partonic channels. This reveals that the total NNLO correction comes from an intricate interplay between all contributing channels, with different ones dominating in different $\eta_{jet}$ regions.  Only the gluon-lepton partonic process is negligible over all of phase space.  For negative $\eta_{jet}$, the dominant contribution is given by the quark-quark process.  As discussed before, this appears first at ${\cal O}(\alpha^2\alpha_s^2)$.  It is therefore effectively treated at leading-order in our calculation, and consequently has a large scale dependence.  We note that the quark-in-lepton distribution from Eq.~(\ref{eq:qinl}) is larger at high-$x$ than the corresponding photon-in-lepton one, leading to this channel being larger in the negative $\eta_{jet}$ region.  At high $\eta_{jet}$, the distribution receives sizable contributions from the gluon-photon process.  No single partonic channel furnishes a good approximation to the shape of the full NNLO correction.

\section{Conclusions} \label{sec:conc}

We have presented in this paper the full calculation of the ${\cal O}(\alpha^2 \alpha_s^2)$ perturbative corrections to jet production in electron-nucleus collisions.  To obtain this result we have utilized the $N$-jettiness subtraction scheme introduced to allow NNLO calculations in hadronic collisions.   We have described the application of this method to inclusive jet production in detail, and have shown that upon integration over the final-state hadronic phase that we reproduce the known NNLO result for the inclusive structure functions.  Our results have been implemented in a numerical program {\tt DISTRESS} that we plan to make publicly available for future phenomenological studies.

We have shown numerical results for jet production at a proposed future EIC. Several new partonic channels appear at the ${\cal O}(\alpha^2 \alpha_s^2)$ level, which have an important effect on the kinematic distributions of the jet.  No single partonic channel furnishes a good approximation to the full NNLO result.  The magnitude of the corrections we find indicate that higher-order predictions will be an important part of achieving the precision understanding of proton structure desired at the EIC, and we expect that the methods described here will be an integral part of achieving this goal.

\vspace{0.5cm}
\noindent
\mysection{Acknowledgements}

G.~A. is supported by the NSF grant PHY-1520916.  R.~B. is supported by the DOE contract DE-AC02-06CH11357. X.~L. is supported by the DOE grant DE-FG02-93ER-40762.  F.~P. is supported by the DOE grants DE-FG02-91ER40684 and DE-AC02-06CH11357.  This research used resources of the National Energy Research Scientific Computing Center, a DOE Office of Science User Facility supported by the Office of Science of the U.S. Department of Energy under Contract No. DE-AC02-05CH11231.  It also used resources of the Argonne Leadership Computing Facility, which is a DOE Office of Science User Facility supported under Contract DE-AC02-06CH11357.  This research was supported in part by the NSF under Grant No. NSF PHY11-25915 to the Kavli Institute of Theoretical Physics in Santa Barbara, which we thank for hospitality during the completion of this manuscript.


\begin{thebibliography}{99}

\bibitem{Kang:2011jw} 
  Z.~B.~Kang, A.~Metz, J.~W.~Qiu and J.~Zhou,
  Phys.\ Rev.\ D {\bf 84}, 034046 (2011), 
  [arXiv:1106.3514 [hep-ph]].
     
\bibitem{Kang:2012zr} 
  Z.~B.~Kang, S.~Mantry and J.~W.~Qiu,
  Phys.\ Rev.\ D {\bf 86}, 114011 (2012),
  [arXiv:1204.5469 [hep-ph]].
  
\bibitem{deFlorian:2008mr} 
  D.~de Florian, R.~Sassot, M.~Stratmann and W.~Vogelsang,
  Phys.\ Rev.\ Lett.\  {\bf 101}, 072001 (2008),
  [arXiv:0804.0422 [hep-ph]].
  
  
 \bibitem{Hinderer:2015hra} 
  P.~Hinderer, M.~Schlegel and W.~Vogelsang,
  Phys.\ Rev.\ D {\bf 92}, no. 1, 014001 (2015),
  [arXiv:1505.06415 [hep-ph]].
 
 \bibitem{Gamberg:2014eia} 
  L.~Gamberg, Z.~B.~Kang, A.~Metz, D.~Pitonyak and A.~Prokudin,
  Phys.\ Rev.\ D {\bf 90}, no. 7, 074012 (2014),
  [arXiv:1407.5078 [hep-ph]].
 
 \bibitem{Boughezal:2015dva} 
  R.~Boughezal, C.~Focke, X.~Liu and F.~Petriello,
  Phys.\ Rev.\ Lett.\  {\bf 115}, no. 6, 062002 (2015),
  [arXiv:1504.02131 [hep-ph]].
 
 \bibitem{Gaunt:2015pea}   
  J.~Gaunt, M.~Stahlhofen, F.~J.~Tackmann and J.~R.~Walsh,
  JHEP {\bf 1509}, 058 (2015)
  [arXiv:1505.04794 [hep-ph]].
   
  \bibitem{Boughezal:2015aha} 
  R.~Boughezal, C.~Focke, W.~Giele, X.~Liu and F.~Petriello,
  Phys.\ Lett.\ B {\bf 748}, 5 (2015),
  [arXiv:1505.03893 [hep-ph]].
 
 \bibitem{Boughezal:2015ded} 
 R.~Boughezal, J.~M.~Campbell, R.~K.~Ellis, C.~Focke, W.~T.~Giele, X.~Liu and F.~Petriello,
  Phys.\ Rev.\ Lett.\  {\bf 116}, no. 15, 152001 (2016),
  [arXiv:1512.01291 [hep-ph]].
   
 \bibitem{Boughezal:2016dtm} 
  R.~Boughezal, X.~Liu and F.~Petriello,
  arXiv:1602.06965 [hep-ph].

\bibitem{Boughezal:2016isb} 
  R.~Boughezal, X.~Liu and F.~Petriello,
  arXiv:1602.08140 [hep-ph].
 
\bibitem{Boughezal:2016yfp} 
  R.~Boughezal, X.~Liu and F.~Petriello,
  Phys.\ Lett.\ B {\bf 760}, 6 (2016),
  [arXiv:1602.05612 [hep-ph]].

 \bibitem{Berger:2016inr} 
  E.~L.~Berger, J.~Gao, C.~S.~Li, Z.~L.~Liu and H.~X.~Zhu,
  Phys.\ Rev.\ Lett.\  {\bf 116}, no. 21, 212002 (2016),
  [arXiv:1601.05430 [hep-ph]].
  
  \bibitem{Currie:2016ytq} 
  J.~Currie, T.~Gehrmann and J.~Niehues,
  arXiv:1606.03991 [hep-ph].
 
\bibitem{Kosower:1997zr} 
  D.~A.~Kosower,
  Phys.\ Rev.\ D {\bf 57}, 5410 (1998),
  doi:10.1103/PhysRevD.57.5410.

\bibitem{WW}
C. F. von Weizs\"{a}cker, Z. Phys. {\bf 88}, 612 (1934);
E. J. Williams, Phys. Rev. {\bf 45}, 729 (1934).    

\bibitem{Stewart:2010tn} 
  I.~W.~Stewart, F.~J.~Tackmann and W.~J.~Waalewijn,
  Phys.\ Rev.\ Lett.\  {\bf 105}, 092002 (2010)
  [arXiv:1004.2489 [hep-ph]].
  
 \bibitem{Kang:2013wca} 
  Z.~B.~Kang, X.~Liu, S.~Mantry and J.~W.~Qiu,
  Phys.\ Rev.\ D {\bf 88}, 074020 (2013)
  [arXiv:1303.3063 [hep-ph]].
 
  
\bibitem{Kang:2013nha} 
  D.~Kang, C.~Lee and I.~W.~Stewart,
  Phys.\ Rev.\ D {\bf 88}, 054004 (2013)
  [arXiv:1303.6952 [hep-ph]].
        
    
\bibitem{Stewart:2009yx} 
  I.~W.~Stewart, F.~J.~Tackmann and W.~J.~Waalewijn,
  Phys.\ Rev.\ D {\bf 81}, 094035 (2010).

\bibitem{Stewart:2010pd} 
  I.~W.~Stewart, F.~J.~Tackmann and W.~J.~Waalewijn,
  Phys.\ Rev.\ Lett.\  {\bf 106}, 032001 (2011).
      
    
    
\bibitem{Daleo:2009yj} 
  A.~Daleo, A.~Gehrmann-De Ridder, T.~Gehrmann and G.~Luisoni,
  JHEP {\bf 1001}, 118 (2010),
   [arXiv:0912.0374 [hep-ph]].
    
\bibitem{Boughezal:2010mc} 
  R.~Boughezal, A.~Gehrmann-De Ridder and M.~Ritzmann,
  JHEP {\bf 1102}, 098 (2011),
  [arXiv:1011.6631 [hep-ph]].
    
\bibitem{GehrmannDeRidder:2012ja} 
  A.~Gehrmann-De Ridder, T.~Gehrmann and M.~Ritzmann,
  JHEP {\bf 1210}, 047 (2012),
  [arXiv:1207.5779 [hep-ph]].
    
    
\bibitem{Catani:1996vz} 
  S.~Catani and M.~H.~Seymour,
  Nucl.\ Phys.\ B {\bf 485}, 291 (1997)
  Erratum: [Nucl.\ Phys.\ B {\bf 510}, 503 (1998)],
  [hep-ph/9605323].
    
        
\bibitem{Bauer:2000ew} 
  C.~W.~Bauer, S.~Fleming and M.~E.~Luke,
  Phys.\ Rev.\ D {\bf 63}, 014006 (2000);
 C.~W. Bauer, S.~Fleming, D.~Pirjol, and I.~W. Stewart,
\newblock Phys. Rev. {\bf D63}, 114020 (2001), hep-ph/0011336;
  C.~W.~Bauer and I.~W.~Stewart,
  Phys.\ Lett.\ B {\bf 516}, 134 (2001)
  [hep-ph/0107001];
C.~W. Bauer, D.~Pirjol, and I.~W. Stewart,
\newblock Phys. Rev. {\bf D65}, 054022 (2002), hep-ph/0109045;
C.~W. Bauer, S.~Fleming, D.~Pirjol, I.~Z. Rothstein, and I.~W. Stewart,
\newblock Phys. Rev. {\bf D66}, 014017 (2002), hep-ph/0202088.

\bibitem{Zijlstra:1992qd} 
  E.~B.~Zijlstra and W.~L.~van Neerven,
  Nucl.\ Phys.\ B {\bf 383}, 525 (1992).

\bibitem{Becher:2006qw} 
  T.~Becher and M.~Neubert,
  Phys.\ Lett.\ B {\bf 637}, 251 (2006),
  [hep-ph/0603140].

\bibitem{Gaunt:2014xga} 
  J.~R.~Gaunt, M.~Stahlhofen and F.~J.~Tackmann,
  JHEP {\bf 1404}, 113 (2014),
  [arXiv:1401.5478 [hep-ph]].
  
  \bibitem{Boughezal:2015eha} 
  R.~Boughezal, X.~Liu and F.~Petriello,
  Phys.\ Rev.\ D {\bf 91}, no. 9, 094035 (2015),
  [arXiv:1504.02540 [hep-ph]].

\bibitem{Dulat:2015mca} 
  S.~Dulat {\it et al.},
  Phys.\ Rev.\ D {\bf 93}, no. 3, 033006 (2016),
  [arXiv:1506.07443 [hep-ph]].
  
\bibitem{Cacciari:2008gp} 
  M.~Cacciari, G.~P.~Salam and G.~Soyez,
  JHEP {\bf 0804}, 063 (2008),
  [arXiv:0802.1189 [hep-ph]].
  

  
\end{thebibliography}
\end{document}